\documentclass{article}
\usepackage{amsmath}
\usepackage{amsfonts}
\usepackage{amssymb}
\usepackage{graphicx}

\input{tcilatex}

\begin{document}

\title{Information and Entropy\thanks{%
Presented at MaxEnt 2007, the 27th International Workshop on Bayesian
Inference and Maximum Entropy Methods (July 8-13, 2007, Saratoga Springs,
New York, USA).}}
\author{Ariel Caticha \\
{\small Department of Physics, University at Albany-SUNY, }\\
{\small Albany, NY 12222, USA.}}
\date{}
\maketitle

\begin{abstract}
What is information? Is it physical? We argue that in a Bayesian theory the
notion of information must be defined in terms of its effects on the beliefs
of rational agents. Information is whatever constrains rational beliefs and
therefore it is the force that induces us to change our minds. This problem
of updating from a prior to a posterior probability distribution is tackled
through an eliminative induction process that singles out the logarithmic
relative entropy as the unique tool for inference. The resulting method of
Maximum relative Entropy (ME), which is designed for updating from \emph{%
arbitrary} priors given information in the form of \emph{arbitrary}
constraints, includes as special cases both MaxEnt (which allows arbitrary
constraints) and Bayes' rule (which allows arbitrary priors). Thus, ME
unifies the two themes of these workshops -- the Maximum Entropy and the
Bayesian methods -- into a single general inference scheme that allows us to
handle problems that lie beyond the reach of either of the two methods
separately. I conclude with a couple of simple illustrative examples.
\end{abstract}

\section{Introduction}

The general problem of inductive inference is to update from a prior
probability distribution to a posterior distribution when new information
becomes available. This raises several basic questions which are the subject
of this paper. First, what is information? It is clear that data
\textquotedblleft contains\textquotedblright\ or \textquotedblleft
conveys\textquotedblright\ information, but what does this precisely mean?
Is information some sort of physical fluid that can be contained or
transported? Is information \emph{physical}? Can we measure amounts of
information? Do we need to? What is entropy?

A second set of questions revolves around our methods to process
information. We know that Bayes' rule is the natural way to update
probabilities when the new information is in the form of data and we know
that Jaynes' method of maximum entropy, MaxEnt, is designed to handle
information in the form of constraints \cite{Jaynes57}. At first sight these
two methods appear unrelated. Are they compatible with each other? Are there
other methods? Moreover, the range of applicability of either method is
somewhat limited: Bayes' rule can handle arbitrary priors and data, and it
can even handle some constraints, but not arbitrary constraints. On the
other hand, MaxEnt can handle arbitrary constraints even data, but not
arbitrary priors. Can we extend these methods?

As discussed in \cite{Caticha03} the Shannon-Jaynes interpretation of
entropy as a measure of uncertainty or of amount of information is somewhat
problematic. The issue is not purely academic because the way equations are
set up to solve a problem and even the kind of problems that we are willing
to consider are affected by the particular meaning attributed to quantities
such as entropy or probability. The Shannon-Jaynes interpretation was fairly
adequate for their purposes, namely, communication theory and statistical
mechanics, but it is not at all clear that their entropy with its attendant
interpretation was the appropriate tool for the very different problem of
updating probabilities.

The important contribution of Shore and Johnson \cite{ShoreJohnson80} was
the realization that any confusion surrounding the meaning of entropy could
be, if not resolved, at least evaded by directly axiomatizing the procedure
for updating probabilities instead of seeking dubious measures for a vaguely
defined notion of information. Their argument, which is based on demanding
consistency -- if a problem can be solved in two different ways the two
solutions must agree -- is fundamentally sound. However, the detailed
assumptions in their derivation have been criticized in \cite{Karbelkar86,
Uffink95}.

Another approach to entropy was proposed by Skilling \cite{Skilling88}.
Although his axioms were clearly inspired by Shore and Johnson, the method
was very different in two respects. First, Skilling was not directly
concerned with the problem of updating probabilities; his method was
designed for the determination of positive-additive functions such as
intensities in an image. In retrospect we see that the application to this
particular problem was quite unfortunate because when the method failed to
produce good image reconstructions the natural reaction was a widespread
loss of confidence about entropy methods in general.

The second difference, which I think is a truly significant contribution, is
that Skilling's approach is a systematic method for induction. He spelled
out in full detail how to construct a general theory from known special
cases. The fundamental inductive principle is deceptively trivial: \emph{`If
a general theory exists it must apply to special cases'.} The basic idea is
that when there exists a special case that happens to be known all candidate
theories that fail to reproduce it must be discarded. Thus, the known
special cases -- called the axioms of the theory -- constrain the form of
the general theory, and the idea is that a sufficient number of such
constraints will determine the general theory completely. Of course, there
is always the unfortunate possibility that the desired general theory does
not exist, but if it does, then the search can be conducted in a systematic
and orderly way.

Philosophers already had a name for such a method: they called it \emph{%
eliminative induction} \cite{Earman92}. On the negative side, eliminative
induction, like any other form of induction, is not guaranteed to work. It
failed, for example, in Skilling's image reconstruction problem. On the
positive side, eliminative induction adds an interesting twist to Popper's
scientific methodology. According to Popper scientific theories can never be
proved right, they can only be proved false; a theory is corroborated only
to the extent that all attempts at falsifying it have failed. Eliminative
induction is fully compatible with Popper's notions but the point of view is
just the opposite. Instead of focusing on \emph{failure} to falsify one
focuses on \emph{success}: it is the successful falsification of all rival
theories that corroborates the surviving one. The advantage is that one
acquires a more explicit understanding of why competing theories are
eliminated.

The present paper is the third in a sequence devoted to clarifying the use
of relative entropy as a tool for processing information and updating
probabilities \cite{Caticha03, CatichaGiffin06}. In \cite{Caticha03} we
applied Skilling's method to the problem of Shore and Johnson. The answer to
the question `What is entropy?' turns out to be trivial and somewhat
surprising: \emph{entropy needs no interpretation}. We do not need to know
what `entropy' means, we only need to know how to use it. This explains why
the \textquotedblleft correct\textquotedblright\ interpretation had been so
elusive -- there is none. In \cite{Caticha03} and then again in \cite%
{CatichaGiffin06} the special cases, the axioms, were increasingly polished
to clarify how alternative entropies are ruled out. Furthermore, in \cite%
{Caticha03} we also discussed the question, central to any general method of
updating, of the extent to which the distribution of maximum entropy is to
be preferred over all others, the extent to which distributions with
entropies less than the maximum are to be ruled out.

In this paper we review how eliminative induction leads to a unique
candidate for a general theory of inference, the method of Maximum relative
Entropy (ME), which is designed for updating from \emph{arbitrary} priors
given information in the form of \emph{arbitrary} constraints. The three
axioms used in \cite{CatichaGiffin06} -- locality, coordinate invariance,
and consistency for independent subsystems -- are sufficient to single out
the logarithmic relative entropy as the unique tool for updating. In
particular, we wish to elaborate further on the use of the third axiom --
consistency for independent subsystems -- to eliminate alternative entropies 
\cite{Renyi61}.

The idea is rather simple. The known special cases covered under axiom 3
also include situations in which we have a large number $N$ of independent
identical systems where all sorts of inferences can be reliably carried out
using various asymptotic techniques (laws of large numbers, large deviation
theory, etc.). The close connection with the method of maximum entropy has
been repeatedly emphasized by several authors \cite{vanCampenhoutCover81}-%
\cite{Grendar01}. We conclude that the logarithmic relative entropy is the
only candidate for a general method for updating probabilities. Alternative
entropies can be useful for other purposes -- for example, when studying the
information geometry of statistical manifolds -- but not for a general
theory of updating.

In \cite{CatichaGiffin06} we showed that the ME method includes both MaxEnt
and Bayes' rule as special cases and therefore it unifies the two dominant
themes of these workshops -- the Maximum Entropy and Bayesian methods --
into a single general inference scheme that allows us to handle problems
that lie beyond the reach of either of the two methods separately. I
conclude with a couple of simple illustrative examples.

In a companion paper \cite{GiffinCaticha07} we discuss the problem of
multiple constraints. Should the constraints be processed simultaneously or
sequentially and, if so, in what order? There we also give an explicit
example in which ME is used to simultaneously process information in the
form of data and moment constraints.

\section{What is information?}

It is not unusual these days to hear that systems \textquotedblleft
carry\textquotedblright\ or \textquotedblleft contain\textquotedblright\
information and that \textquotedblleft information is
physical\textquotedblright . This mode of expression can perhaps be traced
to the origins of information theory in Shannon's theory of communication.
We say that we have received information when among the vast variety of
messages that could conceivably have been generated by a distant source, we
discover which particular message was actually sent. It is thus that the
message \textquotedblleft carries\textquotedblright\ information. The
analogy with physics is straightforward: the set of all possible states of a
physical system can be likened to the set of all possible messages, and the
actual state of the system corresponds to the message that was actually
sent. Thus, the system \textquotedblleft conveys\textquotedblright\ a
message: the system \textquotedblleft carries\textquotedblright\ information
about its own state. Sometimes the message might be difficult to read, but
it is there nonetheless.

This language -- information is physical\ -- useful as it has turned out to
be, does not exhaust the meaning of the word `information'. The goal of
information theory, or better, communication theory, is to characterize the
sources of information, to measure the capacity of the communication
channels, and to learn how to control the degrading effects of noise. It is
somewhat ironic but nevertheless true that this \textquotedblleft
information\textquotedblright\ theory\ is unconcerned with the central
Bayesian issue of how the message affects the beliefs of a rational agent. A
fully Bayesian information theory demands an explicit account of the
relation between information and beliefs.

Our desire to update from one state of belief to another is driven by the
conviction that not all probability assignments are equally good. One can
argue that what makes one probability assignment better than another is that
it better reflects some objective\ feature of the world, that it provides a
better guide to the \textquotedblleft truth\textquotedblright\ -- whatever
this might mean. The updating mechanism is supposed to allow us to
incorporate information about the world into our beliefs.

The implication is that when confronted with new information\ our choices as
to what we are honestly and rationally allowed to believe should become
correspondingly restricted. This, I propose, is the defining characteristic
of information: \emph{Information is whatever constrains rational beliefs}.
An important aspect of this notion is that for a rational agent the updating
is not optional; it is a moral imperative. \emph{Information is whatever
forces\ a change of rational beliefs. }

Our definition captures an idea of information that is directly related to
changing our minds: information is the driving force behind the process of
learning. Note also that although there is no need to talk about amounts of
information, whether measured in units of bits or otherwise, our notion of
information allows precise quantitative calculations. Indeed, by information
in its most general form, we mean the set of constraints on the family of
acceptable posterior distributions and this is precisely the kind of
information the method of maximum entropy has been designed to handle.

It may be worthwhile to point out an analogy with Newtonian dynamics. The
state of motion of a system is described in terms of momentum -- the
\textquotedblleft quantity\textquotedblright\ of motion -- while the change
from one state to another is explained in terms of an applied force.
Similarly, in Bayesian inference a state of belief is described in terms of
probabilities -- the \textquotedblleft quantity\textquotedblright\ of belief
-- and the change from one state to another is due to information. Just as a
force is defined as that which induces a change in motion, so information is
that which induces a change of beliefs.

\section{Updating probabilities: the ME method}

Consider a variable $x$ which can be discrete or continuous, in one or
several dimensions. The uncertainty about $x$ is described by a probability
distribution $q(x)$. Our goal is to update from the prior distribution $q(x)$
to a posterior distribution $P(x)$ when new information -- that is,
constraints -- becomes available. The constraints could be given in terms of
expected values but this is not necessary. The question is: of all those
distributions $p(x)$ within the family defined by the constraints, which do
we select?

As suggested by Skilling \cite{Skilling88} to select the posterior it seems
reasonable to rank the candidate distributions in \emph{order of increasing
preference}. It is clear that to accomplish this goal the ranking must be
transitive: if distribution $p_{1}$ is preferred over distribution $p_{2}$,
and $p_{2}$ is preferred over $p_{3}$, then $p_{1}$ is preferred over $p_{3}$%
. Such transitive rankings are represented by assigning to each $p(x)$ a
real number $S[p]$, which we will henceforth call entropy, in such a way
that if $p_{1}$ is preferred over $p_{2}$, then $S[p_{1}]>S[p_{2}]$. The
selected distribution $P$ (one or possibly many, for on the basis of the
available information there may be several equally preferred distributions)
will be that which maximizes the entropy $S[p]$. We are thus led to a method
of Maximum Entropy (ME) that is a variational method involving entropies
which are real numbers. These features are imposed on purpose; they are
dictated by the function that the ME method is \emph{designed }to perform.

Next, to define the ranking scheme, we must decide on the functional form of 
$S[p]$. First, the purpose of the method is to update from priors to
posteriors. The ranking scheme must depend on the particular prior $q$ and
therefore the entropy $S$ must be a functional of both $p$ and $q$. Thus the
entropy $S[p,q]$ produces a ranking of the distributions $p$ \emph{relative}
to the given prior $q$: $S[p,q]$ is the entropy of $p$ \emph{relative} to $q$%
. Accordingly $S[p,q]$ is commonly called \emph{relative entropy}. Since all
entropies are relative, even when relative to a uniform distribution, the
modifier `relative' is redundant and will be dropped.

Second, since we deal with incomplete information the method, by its very
nature, cannot be deductive: \emph{the} \emph{method must be inductive}. The
best we can do is use those special cases where we know what the preferred
distribution should be to eliminate those entropy functionals $S[p,q]$ that
fail to provide the right update. The known\ special cases will be called
(perhaps inappropriately) the \emph{axioms} of the theory. They play a
crucial role: they define what makes one distribution preferable over
another.

The three axioms below are chosen to reflect the conviction that information
collected in the past and codified into the prior distribution is very
valuable and should not be frivolously discarded. This attitude is maximally
conservative: the only aspects of one's beliefs that should be updated are
those for which new evidence has been supplied. Furthermore, since the
axioms do not tell us what and how to update, they merely tell us what not
to update, they have the added bonus of maximizing objectivity -- there are
many ways to change something but only one way to keep it the same. Thus, we
adopt the

\textbf{Principle of Minimal Updating} (PMU): \emph{Beliefs should be
updated only to the extent required by the new information.}

\noindent The three axioms, a brief motivation for them, and their
consequences for the functional form of the entropy are listed below; more
details and proofs are given in \cite{Caticha03} and \cite{CatichaGiffin06}.
As will become immediately apparent the axioms do not refer to merely three
cases; any induction from such a weak foundation would hardly be reliable.
The reason the axioms are convincing and so constraining is that they refer
to three infinitely large classes of known special cases.

\textbf{Axiom 1: Locality}. \emph{Local information has local effects.}

\noindent Suppose the information to be processed does not refer to a
particular subdomain $\mathcal{D}$ of the space $\mathcal{X}$ of $x$'s. In
the absence of any new information about $\mathcal{D}$ the PMU demands we do
not change our minds about $\mathcal{D}$. Thus, we design the inference
method so that $q(x|\mathcal{D})$, the prior probability of $x$ conditional
on $x\in \mathcal{D}$, is not updated. The selected conditional posterior is 
$P(x|\mathcal{D})=q(x|\mathcal{D})$. The consequence of axiom 1 is that
non-overlapping domains of $x$ contribute additively to the entropy.
Dropping additive terms and multiplicative factors that do not affect the
overall ranking, the entropy functional can be simplified to the form 
\begin{equation}
S[p,q]=\int dx\,F\left( p(x),q(x),x\right) \ ,  \label{axiom1}
\end{equation}%
where $F$ is some unknown function.

\textbf{Axiom 2: Coordinate invariance.} \emph{The system of coordinates
carries no information. }

\noindent The points $x$ can be labeled using any of a variety of coordinate
systems. One can \emph{always} change coordinates but this should not affect
the ranking of the distributions. The consequence of axiom 2 is that $S[p,q]$
can be written in terms of coordinate invariants such as $dx\,m(x)$ and $%
p(x)/m(x)$, and $q(x)/m(x)$: 
\begin{equation}
S[p,q]=\int dx\,m(x)\Phi \left( \frac{p(x)}{m(x)},\frac{q(x)}{m(x)}\right) ~.
\label{axiom2}
\end{equation}%
(Again, additive terms and multiplicative factors that do not affect the
overall ranking have been dropped.) Thus the unknown function $F$ which had
three arguments has been replaced by two unknown functions, one is a density 
$m(x)$, and the other is a function $\Phi $ with two arguments. Next we
determine the density $m(x)$ by invoking the locality axiom 1 once again.

\textbf{Axiom 1 (special case): }\emph{When there is no new information
there is no reason to change one's mind. }

\noindent When no new information is available the domain $\mathcal{D}$ in
axiom 1 coincides with the whole space $\mathcal{X}$. The conditional
probabilities $q(x|\mathcal{D})=q(x|\mathcal{X})=q(x)$ should not be updated
and the selected posterior distribution coincides with the prior, $P(x)=q(x)$%
. The consequence is that up to normalization $m(x)$ must be the prior
distribution $q(x)$, which restricts the entropy to functionals of the form 
\begin{equation}
S[p,q]=\int dx\,q(x)\Phi \left( \frac{p(x)}{q(x)}\right) ~.
\end{equation}

\textbf{Axiom 3:\ Consistency for independent subsystems}. \emph{When a
system is composed of subsystems that are }known\emph{\ to be independent it
should not matter whether the inference procedure treats them separately or
jointly. }

Suppose the information on two independent subsystems 1 and 2 is such that
the prior distributions $q_{1}(x_{1})$ and $q_{2}(x_{2})$ are respectively
updated to $P_{1}(x_{1})$ and $P_{2}(x_{2})$ when they are treated
separately. When treated as a single system the joint prior is $%
q_{1}(x_{1})q_{2}(x_{2})$ and the family of potential posteriors is $%
p(x_{1},x_{2})=p_{1}(x_{1})p_{2}(x_{2})$. The entropy functional must be
such that the selected posterior is $P_{1}(x_{1})P_{2}(x_{2})$. The
consequence of axiom 3 for this particular case of just two subsystems is
that entropies are restricted to the one-parameter family given by 
\begin{equation}
S_{\eta }[p,q]=\frac{1}{\eta (\eta +1)}\left[ 1-\int dx\,p(x)\left( \frac{%
p(x)}{q(x)}\right) ^{\eta }\right] ~.  \label{S sub eta}
\end{equation}%
Once again, additive terms and multiplicative factors that do not affect the
overall ranking scheme can be freely chosen. The $\eta =0$ case reproduces
the usual logarithmic relative entropy, 
\begin{equation}
S[p,q]=-\int dx\,p(x)\log \frac{p(x)}{q(x)}  \label{S}
\end{equation}%
[Use $y^{\eta }=\exp \eta \log y\approx 1+\eta \log y$ in eq.(\ref{S sub eta}%
) and let $\eta \rightarrow 0$ to get eq.(\ref{S}).]

In \cite{CatichaGiffin06} we argued that the index $\eta $ has to be the
same for all systems. To see why consider any two independent systems
characterized by $\eta _{1}$ and $\eta _{2}$. Consistency between the joint
and separate updates requires that $\eta _{1}=\eta _{2}$ therefore $\eta $
must be a universal constant. From the success of statistical mechanics as a
theory of inference we inferred that the value of this constant must be $%
\eta =0$ leading to the logarithmic entropy, eq.(\ref{S}). Here we offer a
different argument also based on a broader application of axiom 3:

\textbf{Axiom 3 (special case):\ Consistency for large numbers of
independent identical subsystems}.

\noindent The known special cases covered under axiom 3 include situations
in which we have a large number $N$ of independent identical systems. In
such cases either the weak law of large numbers or large deviation theory in
the form of Sanov's theorem are sufficient to make the desired inferences.
Entropy considerations are not needed.

Let the $x$ variables be discrete $x_{i}$ with $i=1\ldots m$. The identical
priors for the individual systems are $q_{i}$ and the available information
is that the potential posteriors $p_{i}$ are subject, for example, to an
expectation value constraint such as $\langle a\rangle =A$, where $A$ is
some specified value and $\langle a\rangle =\tsum a_{i}p_{i}$.

Consider the set of $N$ systems treated jointly. Let the number of systems
found in state $i$ be $n_{i}$, and let $f_{i}=n_{i}/N$ be the corresponding
frequency. In the limit of large $N$ the frequencies $f_{i}$ converge (in
probability) to the desired posterior $P_{i}$ while the sample average $\bar{%
a}=\tsum a_{i}f_{i}$ converges (also in probability) to the expected value $%
\langle a\rangle =A$. The probability of a particular frequency distribution 
$f=\{f_{1}\ldots f_{n}\}$ generated by the prior $q$ is multinomial, 
\begin{equation}
Q_{N}\left( f|q\right) =\frac{N!}{n_{1}!\ldots n_{m}!}q_{1}^{n_{1}}\ldots
q_{m}^{n_{m}}\quad \text{with}\quad \tsum\limits_{i=1}^{m}n_{i}=N~,
\end{equation}%
and for large $N$ we have 
\begin{equation}
Q_{N}\left( f|q\right) \approx \exp N(S[f,q]+r_{N})~,
\end{equation}%
where $S[f,q]$ given by eq.(\ref{S}), and where $r_{N}$ is a correction that
vanishes as $N\rightarrow \infty $. To find the most probable frequency
distribution satisfying the constraint $\bar{a}=A$ one maximizes $%
Q_{N}\left( f|q\right) $ subject to $\bar{a}=A$, which is equivalent to
maximizing the entropy $S[f,q]$ subject to $\bar{a}=A$. The corresponding
problem for the individual systems is that of maximizing $S_{\eta }[p,q]$
subject to $\langle a\rangle =A$. The two procedures agree only when we
choose $\eta =0$. Therefore, entropies $S_{\eta }$ with $\eta \neq 0$ are
not consistent with the laws of large numbers and must be discarded.

Csiszar \cite{Csiszar84} and Grendar \cite{Grendar01} have argued that the
asymptotic argument above provides a valid justification for the ME method
of updating. An agent whose prior is $q$ receives the information $%
\left\langle a\right\rangle =A$ which can be reasonably interpreted as a
sample average $\bar{a}=A$ over a large ensemble of $N$ trials. The agent's
beliefs are updated so that the posterior $P$ coincides with the most
probable $f$ distribution. This is quite compelling but, of course, as a
justification of the ME method it is restricted to situations where it is
natural to think in terms of ensembles with large $N$. This justification is
not nearly as compelling for singular events for which large ensembles
either do not exist or are too unnatural and contrived. From our point of
view the asymptotic argument above does not by itself provide a fully
convincing justification for the universal validity of the ME method but it
does provide considerable inductive support. It serves as a valuable
consistency check that must be passed by any inductive inference procedure
that claims to be of \emph{general} applicability.

The results are summarized as follows:

\noindent \textbf{The ME method:} \emph{The objective is to update from a
prior distribution }$q$\emph{\ to a posterior distribution given the
information that the posterior lies within a certain family of distributions 
}$p$\emph{. The selected posterior }$P(x)$\emph{\ is that which maximizes
the entropy }$S[p,q]$\emph{. Since prior information is valuable the
functional }$S[p,q]$\emph{\ has been chosen so that beliefs are updated only
to the extent required by the new information. No interpretation for }$%
S[p,q] $\emph{\ is given and none is needed.}

\section{Bayes' rule and its generalizations}

The problem is to update our beliefs about $\theta \in \Theta $ ($\theta $
represents one or many parameters) on the basis of three pieces of
information: (1) the prior information codified into a prior distribution $%
q(\theta )$; (2) the data $x\in \mathcal{X}$ (obtained in one or many
experiments); and (3) the known relation between $\theta $ and $x$ given by
the model as defined by the sampling distribution or likelihood, $q(x|\theta
)$. The updating consists of replacing the \emph{prior} probability
distribution $q(\theta )$ by a \emph{posterior} distribution $P(\theta )$
that applies after the data has been processed.

The crucial element that will allow Bayes' rule to be smoothly incorporated
into the ME scheme is the realization that before the data information is
available not only we do not know $\theta $, we do not know $x$ either.
Thus, the relevant space for inference is not $\Theta $ but the product
space $\Theta \times \mathcal{X}$ and the relevant joint prior is $%
q(x,\theta )=q(\theta )q(x|\theta )$. We should emphasize that the
information about how $x$ is related to $\theta $ is contained in the \emph{%
functional form} of the distribution $q(x|\theta )$ -- for example, whether
it is a Gaussian or a Cauchy distribution -- and not in the actual values of
the arguments $x$ and $\theta $ which are, at this point, still unknown.

Next we collect data and the observed values turn out to be $x^{\prime }$.
We must update to a posterior that lies within the family of distributions $%
p(x,\theta )$ that reflect the fact that $x$ is known, 
\begin{equation}
p(x)=\tint d\theta \,p(\theta ,x)=\delta (x-x^{\prime })~.
\label{data constraint a}
\end{equation}%
This data information constrains but is not sufficient to determine the
joint distribution 
\begin{equation}
p(x,\theta )=p(x)p(\theta |x)=\delta (x-x^{\prime })p(\theta |x^{\prime })~.
\end{equation}%
Any choice of $p(\theta |x^{\prime })$ is in principle possible. Additional
input is needed and it is at this point that we invoke the Principle of
Minimal Updating: beliefs need to be revised only to the extent required by
the data. Accordingly the conditional prior $q(\theta |x^{\prime })$
requires no revision and the selected posterior $P(x,\theta )$ is such that $%
P(\theta |x^{\prime })=q(\theta |x^{\prime })$, or 
\begin{equation}
P(x,\theta )=\delta (x-x^{\prime })q(\theta |x^{\prime })~.
\end{equation}%
The corresponding marginal posterior probability $P(\theta )$ is 
\begin{equation}
P(\theta )=\tint dx\,P(\theta ,x)=q(\theta |x^{\prime })=q(\theta )\frac{%
q(x^{\prime }|\theta )}{q(x^{\prime })}~,  \label{Bayes rule}
\end{equation}%
which is recognized as Bayes' rule. This is extremely reasonable: we \emph{%
maintain} those beliefs about $\theta $ that are consistent with the data
values $x^{\prime }$ that turned out to be true. Data values that were not
observed are discarded because they are now known to be false. `Maintain' is
the key word: it reflects the PMU in action.

\noindent \textbf{Remark:} Bayes' rule is usually written in the form 
\begin{equation}
q(\theta |x^{\prime })=q(\theta )\frac{q(x^{\prime }|\theta )}{q(x^{\prime })%
}~,
\end{equation}%
and called Bayes' theorem. This formula is very simple; perhaps it is too
simple. It is just a restatement of the product rule -- valid for any $%
x^{\prime }$ whether observed or not -- and therefore it is a simple
consequence of the \emph{internal} consistency of the \emph{prior} beliefs.
The drawback of this formula is that the left hand side is not a \emph{%
posterior} but rather a \emph{prior} (conditional) probability; it obscures
the fact that an additional principle -- the PMU -- was needed for updating.

Next we show that Bayes' rule is consistent with, and indeed, is a special
case of the ME method \cite{CatichaGiffin06}. This is not too surprising
given that the ME is also based on the PMU. According to the ME method the
selected joint posterior $P(x,\theta )$ is that which maximizes the entropy,%
\begin{equation}
S[p,q]=-\tint dxd\theta ~p(x,\theta )\log \frac{p(x,\theta )}{q(x,\theta )}%
~,~  \label{entropy}
\end{equation}%
subject to the appropriate constraints. Note that the information in the
data, eq.(\ref{data constraint a}), represents an \emph{infinite} number of
constraints on the family $p(x,\theta )$: for each value of $x$ there is one
constraint and one Lagrange multiplier $\lambda (x)$. Maximizing $S$, (\ref%
{entropy}), subject to (\ref{data constraint a}) and normalization, 
\begin{equation}
\delta \left\{ S+\alpha \left[ \tint dxd\theta ~p(x,\theta )-1\right] +\tint
dx\,\lambda (x)\left[ \tint d\theta ~p(x,\theta )-\delta (x-x^{\prime })%
\right] \right\} =0~,
\end{equation}%
yields the joint posterior, 
\begin{equation}
P(x,\theta )=q(x,\theta )\,\frac{e^{\lambda (x)}}{Z}~,
\end{equation}%
where $Z$ is a normalization constant, and $\lambda (x)$ is determined from (%
\ref{data constraint a}), 
\begin{equation}
\tint d\theta ~q(x,\theta )\frac{\,e^{\lambda (x)}}{Z}=q(x)\frac{%
\,e^{\lambda (x)}}{Z}=\delta (x-x^{\prime })~,
\end{equation}%
so that the joint posterior is%
\begin{equation}
P(x,\theta )=q(x,\theta )\frac{\,\delta (x-x^{\prime })}{q(x)}=\delta
(x-x^{\prime })q(\theta |x)~,
\end{equation}%
from which we recover Bayes' rule, eq.(\ref{Bayes rule}).

I conclude with a couple of very simple examples that show how the ME allows
generalizations of Bayes' rule. The background for these generalized Bayes
problems is the familiar one: We want to make inferences about some
variables $\theta $ on the basis of information about other variables $x$.
As before, the prior information consists of our prior knowledge about $%
\theta $ given by the distribution $q(\theta )$ and the relation between $x$
and $\theta $ is given by the likelihood $q(x|\theta )$; thus, the prior
joint distribution $q(x,\theta )$ is known. But now the information about $x$
is much more limited.

\noindent \textbf{Example 1.--} The data is uncertain: $x$ is not known. The
marginal posterior $p(x)$ is no longer a sharp delta function but some other
known distribution, $p(x)=P_{D}(x)$. This is still an infinite number of
constraints 
\begin{equation}
p(x)=\tint d\theta \,p(\theta ,x)=P_{D}(x)~,  \label{data constraint b}
\end{equation}%
that are easily handled by ME. Maximizing $S$, (\ref{entropy}), subject to (%
\ref{data constraint b}) and normalization, leads to 
\begin{equation}
P(x,\theta )=P_{D}(x)q(\theta |x)~.
\end{equation}%
The corresponding marginal posterior, 
\begin{equation}
P(\theta )=\tint dx\,P_{D}(x)q(\theta |x)=q(\theta )\tint dx\,P_{D}(x)\frac{%
q(x|\theta )}{q(x)}~,  \label{Jeffrey}
\end{equation}%
is known as Jeffrey's rule.

\noindent \textbf{Example 2.--} Now we have even less information: $p(x)$ is
not known. All we know about $p(x)$ is an expected value 
\begin{equation}
\left\langle f\right\rangle =\tint dx\,p(x)f(x)=F~.
\label{data constraint c}
\end{equation}%
Maximizing $S$, (\ref{entropy}), subject to (\ref{data constraint c}) and
normalization, 
\begin{equation}
\delta \left\{ S+\alpha \left[ \tint dxd\theta ~p(x,\theta )-1\right]
+\lambda \tint dxd\theta ~p(x,\theta )f(x)-F\right\} =0~,
\end{equation}%
yields the joint posterior, 
\begin{equation}
P(x,\theta )=q(x,\theta )\,\frac{e^{\lambda f(x)}}{Z}~,
\end{equation}%
where the normalization constant $Z$ and the multiplier $\lambda $ are
obtained from 
\begin{equation}
Z=\tint dx~q(x)e^{\lambda f(x)}\quad \text{and}\quad \frac{d\log Z}{d\lambda 
}=F~.
\end{equation}%
The corresponding marginal posterior is 
\begin{equation}
P(\theta )=q(\theta )\tint dx\,\,\frac{e^{\lambda f(x)}}{Z}q(x|\theta )~.
\label{Bayes ME}
\end{equation}%
The two posteriors (\ref{Jeffrey}) and (\ref{Bayes ME}) are sufficiently
intuitive that one could have written them down directly without deploying
the full machinery of the ME\ method, but they do serve to illustrate the
essential compatibility of Bayesian and Maximum Entropy methods. A less
trivial example is given in \cite{GiffinCaticha07}.

\section{Conclusions}

Any Bayesian account of the notion of information cannot ignore the fact
that Bayesians are concerned with the beliefs of rational agents. The
relation between information and beliefs must be clearly spelled out. The
definition we have proposed -- that information is that which constrains
rational beliefs and therefore forces the agent to change its mind -- is
convenient for two reasons. First, the information/belief relation very
explicit, and second, the definition is ideally suited for quantitative
manipulation using the ME\ method.

The other main conclusion is that the logarithmic relative entropy is the
only candidate for a general method for updating probabilities -- the ME\
method -- which includes MaxEnt and Bayes' rule as special cases; it unifies
them into a single theory of inductive inference.

It is true that there exist many different ways to define measures of
separation, or divergence between distributions and that these
\textquotedblleft entropies\textquotedblright\ can be useful in a wide
variety of ways. In fact, it was precisely this wealth of possibilities that
Shore and Johnson intended to avoid. These other \textquotedblleft
entropies\textquotedblright\ can be useful for other purposes but not for
updating; at least not for an updating theory that strives to achieve
universal applicability. Let us emphasize that the reason the ME\ method
uses the logarithmic entropy as the tool for updating is not that this
entropy has been shown to provide the \emph{correct} measure of distance --
there are many other such measures. We do not even claim that inferences on
the basis of the ME method are guaranteed to be \emph{correct} -- this is
induction; there are no guarantees. It is just that all alternative
entropies are much worse because in known cases they give answers that are
demonstrably wrong.

\noindent \textbf{Acknowledgements:} I would like to acknowledge valuable
discussions with C. Cafaro, N. Caticha, A. Giffin, K. Knuth, and C. Rodr%
\'{\i}guez.

\end{document}